\documentclass[twocolumn,amsmath,amssymb,showpacs,prx,aps,longbibliography,superscriptaddress]{revtex4-1}
\usepackage[bookmarks=false,linkcolor=blue,urlcolor=blue,colorlinks,citecolor=blue]{hyperref}

\pdfoutput=1 

\usepackage{graphicx}  
\usepackage{dcolumn}   
\usepackage{bm}        
\usepackage{amssymb}   

\newcommand{\be}{\begin{equation}}
\newcommand{\ee}{\end{equation}}
\newcommand{\bea}{\begin{eqnarray}}
\newcommand{\eea}{\end{eqnarray}}

\usepackage{amsmath}
\usepackage{amssymb}
\usepackage{xcolor}

\usepackage[english]{babel}
\usepackage{amsmath,amsthm,amssymb}
\usepackage{amsfonts}
\usepackage{amsopn}
\usepackage{float}
\usepackage{url}
\usepackage{graphicx}
\usepackage{dcolumn}
\usepackage{bm}
\usepackage{color}

\usepackage{float}

\newcommand{\bra}[1]{\langle #1|}
\newcommand{\ket}[1]{|#1\rangle}

\newcommand{\bpm}{\begin{pmatrix}}
\newcommand{\epm}{\end{pmatrix}}
\newcommand{\bmm}{\begin{matrix}}
\newcommand{\emm}{\end{matrix}}

\newcommand{\mtwo}[4]{\left( \begin{array}{cc}#1 & #2\\ #3 & #4 \end{array} \right)}
\newcommand{\vvectwo}[2]{\left( \begin{array}{c}#1 \\ #2 \end{array} \right)}

\begin{document}
\author{Maciej Koch-Janusz}
\affiliation{Institute for Theoretical Physics, ETH Zurich, 8093 Zurich, Switzerland}
\author{Kusum Dhochak}
\affiliation{International Centre for Theoretical Sciences, Tata Institute of Fundamental Research, Bengaluru-560089, India.}
\author{Erez Berg}
\affiliation{Department of Condensed Matter Physics, Weizmann Institute of Science, Rehovot IL-76100, Israel}	
	
\title{Edge--Entanglement correspondence for gapped topological phases with symmetry}

\begin{abstract}
	The correspondence between the edge theory and the entanglement spectrum is firmly established for the chiral topological phases. We study gapped, topologically ordered, non-chiral states with a conserved $U(1)$ charge and show that the entanglement Hamiltonian contains not only the information about topologically distinct edges such phases may admit, but also which  of them will be realized in the presence of symmetry breaking/conserving perturbations. We introduce an exactly solvable, charge conserving lattice model of a $\mathbb{Z}_2$ spin liquid and derive its edge theory and the entanglement Hamiltonian, also in the presence of perturbations. We construct a field theory of the edge and study its RG flow. We show the precise extent of the correspondence between the information contained in the entanglement Hamiltonian and the edge theory.
\end{abstract}

\maketitle

\section{Introduction}
One of the remarkable properties of topological states of matter is the intimate relationship between the physics of the bulk and of the surface.
Customarily referred to as the bulk-edge correspondence, this relationship severely restricts the possible edge theories to those compatible with the bulk order. Conversely, the bulk topological properties, such as the charge of the elementary excitations, can be revealed in measurements on the surface~\cite{PhysRevLett.48.1559,PhysRevLett.79.2526,Heiblum,Dolev}. 

First introduced in the seminal work of X.-G. Wen~\cite{wenbb} in the context
of fractional Quantum Hall (FQHE) states, the correspondence directly linked the K-matrix specifying the Chern-Simons field theory of the gapped bulk with a corresponding matrix describing the structure of gapless chiral Luttinger liquids
on the edge. The correspondence is not, however, restricted to chiral topological states: it also applies to symmetry-protected topological states (SPTs)~\cite{Chen1604,PhysRevB.93.205157} and even, in a weaker form, to fully gapped topological orders.
It is well known, for instance, that a $\mathbb{Z}_2$ spin liquid admits exactly two topologically distinct kinds of gapped edges, whose nature is related to properties of bulk spectrum~\cite{1998quant.ph.11052B}.
The latter example demonstrates that the correspondence is in fact one-to-many: a single topologically ordered bulk can, in general, support several distinct phases of the edge \cite{PhysRevX.3.021009,PhysRevB.89.115116,PhysRevB.88.235103}.

If symmetry is present, it can constrain the allowed edge phases \cite{Lu2016,Barkeshli722}. Thus, the edges of a topologically ordered phase with a given transformation law of the elementary quasi-particles under the symmetry (also known as a ``symmetry-enriched topological phase'' \cite{PhysRevLett.105.246809,PhysRevB.83.195139,PhysRevB.86.115131,PhysRevB.87.155115,PhysRevB.87.104406,Fidkowski2015}) is expected to have a generic phase diagram, with several different phases separated by sharp transitions. These considerations have been used to predict dramatic physical effects at the gapped edges of a $\mathbb{Z}_2$ quantum spin liquid with fractionalized spinon and holon excitations~\cite{Barkeshli722}. Arguments for the generic phase diagram of the edge and the effects of symmetry breaking perturbations have been put forth; however, a concrete model that can flesh out the edge effective Hamiltonian and phase diagram has been lacking.

More recently a new avenue for research of topological states has been opened involving the study of their entanglement properties~\cite{PhysRevLett.90.227902,PhysRevLett.92.096402,PhysRevLett.98.060401,PhysRevB.76.125310,PhysRevLett.96.110404,PhysRevLett.96.110405,PhysRevLett.99.220405,PhysRevLett.113.106801,Hamma200522,PhysRevB.85.235151}. 
In particular, the \emph{entanglement spectrum} of topological phases, obtained by bi-partitioning the system and diagonalizing the density matrix of a one of the subsystems, has been shown to contain analogous information to that of the physical edge\cite{PhysRevLett.101.010504}.
This striking observation was proven in general for chiral topological states using the methods of boundary conformal field theory\cite{PhysRevLett.108.196402}. The relation between the entanglement spectrum and the physical edge extends to SPT phases, as well \cite{PhysRevB.81.064439, PhysRevLett.104.130502}. Many works investigating the entanglement spectra of topological phases have followed \cite{PhysRevLett.105.080501,PhysRevLett.107.157001, PhysRevB.86.245310,PhysRevB.84.205136,PhysRevLett.108.227201,PhysRevB.86.045117}.  The precise amount of information about the edges contained in the entanglement spectrum -- and its relation to the properties of the physical edge -- have been subject of some investigations recently \cite{PhysRevLett.113.060501,PhysRevB.91.125119}.

In this paper we study the edge and the entanglement properties of an exactly solvable model for a symmetry-enriched $\mathbb{Z}_2$ quantum spin liquid with a conserved $U(1)$ charge, introduced in Ref.~\cite{Levin:2011hq}. In this model, the excitations that carry a non-trivial gauge charge (``spinons'') also carry a fractional physical charge. We study the phase diagram of the edge, as a function of both charge-conserving and non-conserving perturbations. At the exactly solvable point of the model, the edge spectrum is macroscopically degenerate.  When small, generic perturbations are introduced at the edge, the low energy effective edge Hamiltonian is found to be of the form of a Bose-Hubbard model. Breaking the $U(1)$ symmetry on the edge causes the appearance of additional ``pairing'' terms in the edge Hamiltonian. 
The phase diagram of the edge includes a gapped, symmetry-preserving phase with condensed visons ($m$ phase), a gapped phase with condensed spinons ($e$ phase) which requires an explicit breaking of the $U(1)$ symmetry \footnote{Physically, the conserved $U(1)$ charge can correspond to the total spin along the $z$ direction in an easy-axis quantum spin liquid. Then, realizing the $e$ phase requires an in-plane magnetic field.}, and a gapless, symmetry-preserving phase.

Next, we study the entanglement properties of a cut through the system. The entanglement Hamiltonian, $\hat{H}_{ent}$, is shown to be described in terms of the same set of degrees of freedom as the edge Hamiltonian, $\hat{H}_e$, and is massively degenerate at the exactly solvable point of the bulk. We derive the entanglement Hamiltonian perturbatively for generic, integrability-breaking perturbations in the bulk, using the Schrieffer-Wolff method technique of Ref. \cite{PhysRevB.91.125119}. $\hat{H}_{ent}$ is then of the same Bose-Hubbard form as the edge Hamiltonian, although its parameters are different. Thus, as parameters of the bulk Hamiltonian are varied, the entanglement Hamiltonian has the same global phase diagram as a physical edge.

This case study supports the notion (as was already noted in Ref.~\cite{2015arXiv151002982H}) that the bulk topological order, in conjunction with the symmetry properties of the quasi-particles, dictate the allowed types of phases that can appear on the boundary, although it does not uniquely determine which one is realized at a particular edge. In this sense, the bulk-edge correspondence is a relation between the bulk topological order and the class of possible edge phases (or the global edge phase diagram). This is true for both a physical edge and an entanglement edge; i.e., the \emph{class} of possible phases realized by the entanglement Hamiltonian is the same as that of a physical edge.

The paper is organised as follows: in section \ref{sec:lattice} we introduce the lattice model, in section \ref{sec:edge} we derive the edge Hamiltonian for the perturbed and unperturbed model; we write down a field theory for the edge and analyse its phase diagram. In section \ref{sec:ent} we calculate the entanglement Hamiltonian of the perturbed model and show its relation to the edge Hamiltonian. Finally in section \ref{sec:co} we recapitulate our results and speculate on their general validity beyond solvable models considered thus far in the literature.

\section{The lattice model}\label{sec:lattice}

We begin by introducing an exactly solvable, charge-conserving,  lattice model of  a $\mathbb{Z}_2$ spin liquid. The bosonic lattice model, a special case of the one we considered in  Ref. \onlinecite{Levin:2011hq}, is in fact closely related to the family of toric code-like  Hamiltonians, with an additional conserved $U(1)$ symmetry. It will allow us to \emph{exactly} compute the properties of both the edge and the entanglement spectrum. It will also be a useful starting point for analysis of the perturbations away from the solvable point.

\begin{figure}[h]
\centering
\includegraphics*[width=9cm]{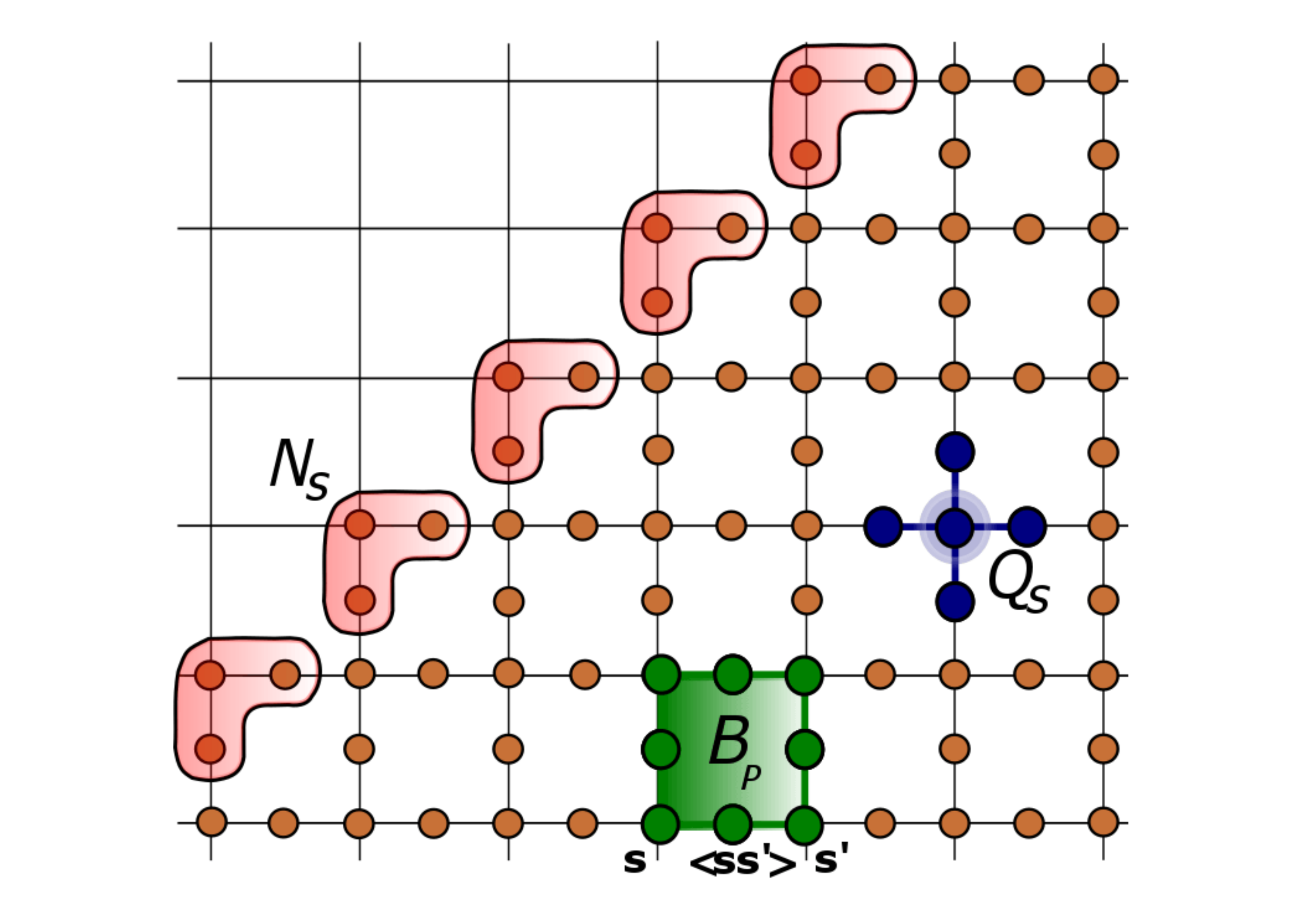}
\caption{Illustration of the lattice model Eq. (\ref{HBose}). The bosons live on both sites $s$ and links $\langle ss' \rangle$ of the lattice. The bulk Hamiltonian $H_b$ is a sum of squares of $Q_s$ terms, which act on a site and adjacent links and the $B_P$ terms, which act on links and sites belonging to the plaquette $P$.}
\label{fig:latticem1}
\end{figure}

\paragraph*{Bulk -- }
The Hilbert space of our model consists of bosonic degrees of freedom which reside on the sites and on the links of the lattice. For the site bosons $n_s$ we use a rotor representation with the creation operator $b^{\dagger}_s = e^{i\theta_s}$ and $[\theta_s,n_s] = i$. The site-boson occupation number can therefore assume any integer (i.e. also negative) value. In contrast, the link variables $n_{ss'}$ are defined as hard-core bosons, i.e. $n_{ss'} \in \{0,1\}$. Mapping the link occupation number $n_{ss'}=0$ to $(0,1)^T$ and $n_{ss'} =1$ to $(1,0)^T$ the creation operator can be written explicitly as:
\begin{equation}\label{cre1} b^{\dagger}_{ss'} = \mtwo{0}{1}{0}{0} \end{equation}

The bulk Hamiltonian $H_b$ can be written as a sum of two terms, one associated with sites $s$, and the other associated with
plaquettes $P$ of a Lieb lattice (see Fig. \ref{fig:latticem1}):
\begin{equation} \label{HBose}
\hat{H}_b = v \sum_s Q_s^2 - \frac{u}{2} \sum_P (B_P + B_P^\dagger)
\end{equation}
We assume $u, v > 0$. The first term is the charging Hamiltonian, which depends on the number of bosons $n_s$, $n_{ss'}$:
\begin{equation}
Q_s = 2n_s + \sum_{s'} n_{ss'}.
\label{Qsop}
\end{equation}
Here, $s'$ are all the neighbors of the site $s$.
The conserved charge (not to be confused with $Q_s$) is simply given by:
\begin{equation}\label{Qel} Q = \sum_{s} n_s + \sum_{\langle ss' \rangle} n_{ss'}, \end{equation}
i.e. the unweighted sum of all site and link boson occupation numbers.

The second term is the hopping Hamiltonian, it can be thought of as a ring exchange term.
It is defined as the product:
\begin{equation}
B_P = U_{12} U_{23} U_{34} U_{41}
\label{Bpop}
\end{equation}
where $U_{ss'}$ is a boson hopping term between the link $\langle ss'\rangle$ and the sites $s$, $s'$:
\begin{equation}
U_{ss'} = b_{ss'} b_{s'}^\dagger  + b_s b_{ss'}^\dagger
\label{Uop}
\end{equation}

The form of this term has a very simple interpretation: it decreases or increases the numbers of bosons on the link modulo $2$ and also makes sure that the charge conservation at the endpoints is obeyed, by hopping the bosons between the links and sites.

The Hamiltonian (\ref{HBose}) is exactly solvable. This is because the operators $Q_s$ and $B_p$ all commute with each other. The first step to prove it is to notice that:
\begin{equation} \label{QOCom}
[Q_t, U_{ss'}] =  ( \delta_{s't}-\delta_{st}) U_{ss'},
\end{equation}
i.e. $U_{ss'}$ decreases $Q_s$ by $1$, increases $Q_{s'}$ by $1$, and leaves $Q_s$ unchanged
at all other sites -- this allows us to think of $U_{ss'}$ as hopping operator for the $Q_s$ charges.
It also follows that $Q_s$ commutes with the product of $U_{ss'}$ around any closed loop, in particular:
\begin{equation}
[Q_s, B_P] = 0.
\end{equation}
Thus, we conclude that $\{Q_s, B_P, B_P^\dagger\}$ all commute, and therefore can be diagonalized simultaneously.
The simultaneous eigenstates of these operators can be labelled by their eigenvalues: $\ket{q_s, b_P}$. The energies are given by:
\begin{equation}
E = v \sum_s q_s^2 - u/2 \sum_P (b_P + b_P^*).
\end{equation}
It is clear from the definition (\ref{Qsop}) that $Q_s$ has integer eigenvalues, so $q_s$ is an integer. We can also show that
\be
B_P^2 = 1, \label{eq:Bp}
\ee
so $b_P$ must be $\pm1$. We conclude that the ground state of $\hat{H}_b$ is the unique state with $q_s = 0$, $b_P = +1$ everywhere.
There are two types of elementary excitations: charge excitations where $q_s = 1$ for some site $s$, and
flux excitations where $b_P = -1$ for some plaquette $P$. Eq. (\ref{QOCom}) shows that the charge excitations are created (and moved) by a string of $U_{ss'}$ operators along a path on the lattice. Analogous operator for the flux excitations is defined by a path on a dual lattice and application of the elementary flux string operator $(-1)^{n_{ss'}}$ on every link $\langle ss' \rangle$ cut by the path. The spectrum of the hopping Hamiltonian, and hence of the whole $\hat{H}_b$ is discrete. In particular, as long as $u, v > 0$, the ground state is gapped.

The $q_s$ charge excitations carry a \emph{fractional} $U(1)$ charge of $\frac{1}{2}$. This is easily seen by computing the difference between expectation values of $n_s$ at site $s$ for states with $q_s =1$ and $q_s=0$. The total number of charge excitations must be even, so that the total charge is integer. This can also be verified explicitly: $Q = \frac{1}{2} \sum_sQ_s$.  The model is in fact topologically ordered: the charges and fluxes exhibit mutual fractional statistics and the ground state is four-fold degenerate on the torus -- for the details we refer to \cite{Levin:2011hq}. The topological order is the same as for a $\mathbb{Z}_2$ gauge theory.

Assuming the system is defined on a cylinder we have also \emph{topological} operators, which commute with $\hat{H}_b$. Their support is any non-contractible loop around the cylinder. Since in this case all non-contractible loops are equivalent to each other modulo contractible ones, there are two such topological operators:
\begin{equation}\label{topop1} T=\prod_{\langle ss' \rangle\in C}U_{ss'} \end{equation}
\begin{equation}\label{topop2} P=\prod_{ \langle ss'\rangle \in C'}(-1)^{n_{ss'}}, \end{equation}
where $C$ is a non-contractible loop on the lattice and $C'$ a non-contractible loop on the dual lattice. The links $\langle ss' \rangle$ in Eq. (\ref{topop2}) are the ones crossed by $C'$. The $T$ operator can be thought of as creating a pair of $Q_s$ quasiparticles, taking one of them around the cylinder, and annihilating them, while the $P$ operator does the same for a $B_P$ vortex. Note that $[T,P]=0$ (on the cylinder). Since both of them have eigenvalues $\pm 1$ there are altogether $4$ topological sectors.
This can be generalized to any nontrivial topology where non-contractible loops exist \footnote{On a cylinder, generically only two of the four states $T=\pm 1$, $P=\pm 1$ are degenerate. Which two are ground states depends on the boundary conditions. On the torus, all four states are degenerate; this can be seen from the fact that there are similar operators defined on the other non-contractible cycle of the torus, which commute with the Hamiltonian and anticommute with $P$,$T$.}

\paragraph*{Edge -- } We consider a system with a ``zigzag'' edge, as depicted in Fig. \ref{fig:latticem1} and we define the edge operators to be all the operators commuting with the bulk Hamiltonian, which are \emph{not} the bulk or the topological operators considered above.

\begin{figure}[h]
\centerline{
\includegraphics*[width=7cm]{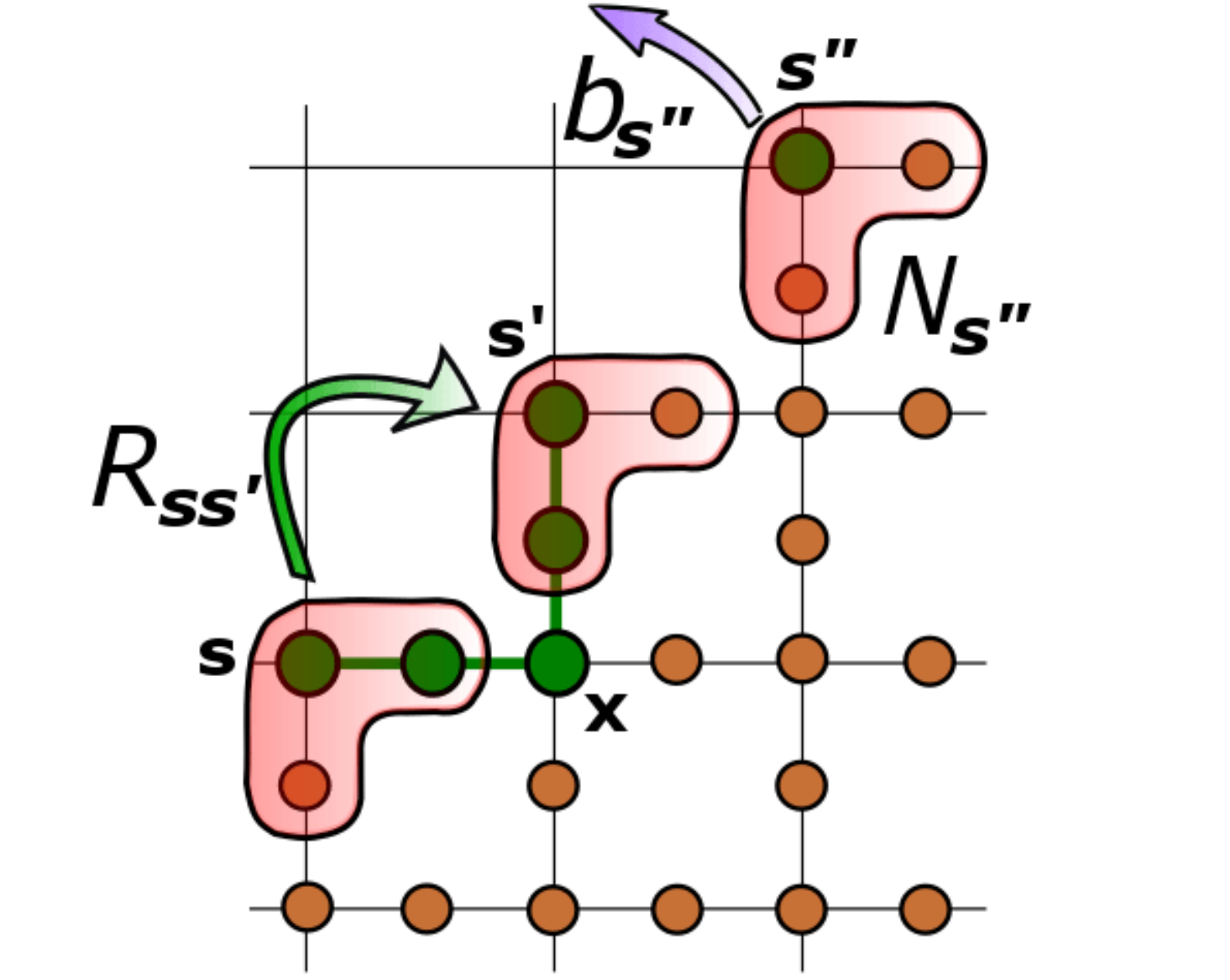}}
\caption{The edge operators commuting with the bulk Hamiltonian. The charging terms $N_s$ are restrictions of the bulk operators $Q_s$. $R_{ss'}$ are the hopping operators for the $N_s$ charges, they are a restriction of the bulk operator $B_P$, i.e. a product of two $U$ operators along the edge. Also the $b_s$ and $b_s^{\dagger}$ operators, which do not conserve the $N_s$ (and the electrical) charge can be thought of as restricted plaquette operator $B_P$. }
\label{fig:latticem2}
\end{figure}

There are three types of such edge operators: the $N_s$ charging operators, depicted in Figs. \ref{fig:latticem1} and \ref{fig:latticem2} in red, are restrictions of the  $Q_s$ operators imposed by the truncated Hilbert space (i.e. $N_s$ is defined exactly like $Q_s$, without the ``missing" links). The $R_{ss'}$, denoted in Fig. \ref{fig:latticem2} with a green arrow, can be thought of as half of a $B_P$ operator, i.e. $R_{ss'} = U_{sx}U_{xs'}$ with $s,s'$ adjacent sites on the edge, and $x$ their common neighbour in the bulk (the sites acted upon by $R_{ss'}$ have been marked green in Fig. \ref{fig:latticem2}). It also acts as a hopping operator for the $N_s$ charges, exactly as $U_{ss'}$ does for $Q_s$:
\begin{equation}\label{eq:rss1}[N_t, R_{ss'}] =  ( \delta_{s't}-\delta_{st}) R_{ss'}, \end{equation}
with $t,s,s'$ belonging to the edge. Finally, there is the $b_s$ operator, with $s$ belonging to edge, denoted by a purple arrow in Fig. \ref{fig:latticem2}. It is a restriction of a plaquette operator, of which only the corner site belongs to the system. Note that $b_s$ and its hermitian conjugate do not conserve the $N_s$ charge (it creates/annihilates \emph{two} units of $N_s$ charge), and hence also the electric charge:
\begin{equation}\label{bs1} [N_t, b_s] = -2\delta_{st}b_s. \end{equation}
Thus, the operator $b_s$ can only appear in the Hamiltonian if charge conservation is broken in the system (at least at the edge). Physically, if the $U(1)$ symmetry of the model is due to conservation of the $z$ component of the total spin in a magnetic system, breaking of the symmetry at the edge can arise at an interface between the spin liquid and an in-plane ferromagnet (in which the symmetry is spontaneously broken).

The edge operators $N_s$, $R_{ss'}$ and $b_s$ do not commute with each other, even though they do commute with the bulk operators (and thus with $\hat{H}_b$). We may define a complete set of commuting observables describing the system with an edge to be the $Q_s$, $B_P$ bulk operators, the $N_s$ operators on the edge, and the topological operator $T$. Note that for non-contractible loops $C,C'$ on lattice/dual lattice defined by the zigzag edge $\partial$, we have:
\begin{equation}\label{topop3} T=\prod_{\langle ss' \rangle\in C}R_{ss'} \end{equation}
\begin{equation}\label{topop4} P=\prod_{ \langle ss'\rangle \in C'}(-1)^{n_{ss'}}=\prod_{s\in \partial} (-1)^{N_s} = (-1)^{\sum_{s\in \partial}N_s} \end{equation}

Eq. (\ref{topop4}) shows the $P$ operator is in fact a \emph{total parity} operator for the $N_s$ bosons. Also, since $P$ is not independent from the full set of $N_s$, we did not include it in the complete set of commuting observables. As we will show below, $\sum_s N_s$ corresponds to the total number of ``spinon'' excitations on the edge. Each such excitation carries a $U(1)$ charge of $1/2$, and is also charged under the emergent gauge field with an ``electric'' charge of $-1$.

\section{The edge Hamiltonian}\label{sec:edge}
Here we write the edge Hamiltonian and map out its phase diagram. We begin with the exactly solvable system and then consider the effect of the perturbations. A field theory for the edge is constructed.
\paragraph*{The solvable edge --}
Consider the system on a semi-infinite cylinder, such that its boundary is the ``zigzag" edge we described (see Fig. \ref{fig:latticem2}). In the absence of any perturbations, the bulk Hamiltonian $\hat{H}_b$ does not contain any of the edge operators, hence the groundstate subspace $\mathcal{H}_0$ in each topological sector consists of an extensive (in the length of the boundary) number of degenerate states, labelled by different eigenvalues of all the $N_s$ operators. Thus the edge spectrum is flat and the edge Hamiltonian $\hat{H}_e$ is identically zero:
\begin{equation}\label{hedge1} \hat{H}_e = 0. \end{equation}

Under the action of Hamiltonian $\hat{H}_b$, the total Hilbert space splits into massively degenerate subspaces, separated by finite gaps:
\begin{equation}\label{hilbertspacedecomp}\mathcal{H} = \bigoplus_{\alpha \geq 0} \mathcal{H}_\alpha.\end{equation}
  The subspace energy is a function of the number of \emph{bulk} quasiparticles. Its degeneracy comes from the edge operators, which do not appear in $\hat{H}_b$, and from the presence of topological sectors. In particular, as we have seen, the degenerate ground state subspace (assuming for concreteness  $u=v=1$) is separated by a gap of 1 from the first excited state with a single quasiparticle.

\paragraph*{The perturbed system --}

Let us now consider adding small perturbations $\epsilon \hat{V}$ to the system; $\hat{V}$ is a sum of local edge and bulk terms. By small we mean perturbations which do not \emph{qualitatively} change  this picture of (nearly) degenerate subspaces separated by finite gaps. We do allow that they impart dispersion on the levels within the subspace, but we assume that the new eigenstates are adiabatically connected to the original, perfectly degenerate, non-perturbed ones.
The edge Hamiltonian $\hat{H}_e$ is then defined as an effective Hamiltonian acting only within lowest-energy subspace of the perturbed Hamiltonian and generating the dispersion within this space. This picture is formalized below.

The edge Hamiltonian can be derived using the ``effective Hamiltonian"  or Schrieffer-Wolff method\cite{PhysRevB.91.125119}, such that it is still block-diagonal in the unperturbed eigenspaces $\mathcal{H}_\alpha$ at the cost of introducing nontrivial matrix elements between the states \emph{within} $\mathcal{H}_\alpha$ (recall, the unperturbed $\hat{H}_e$ was trivial).
The matrix elements of the effective Hamiltonian, are given (to second order) by:
\begin{gather}\label{sw1} \bra{i,\alpha} \hat{H}_{e} \ket{j,\alpha} = E_i^{\alpha} + \epsilon \bra{i\alpha}\hat{V}\ket{j,\alpha} + \mbox{\ \ \ \ \ \ \ }\\ \nonumber
 +\frac{\epsilon^2}{2}\sum_{k,\beta\not= \alpha} \bra{i,\alpha}\hat{V}\ket{k,\beta} \bra{k,\beta}\hat{V}\ket{j,\alpha}\left(\frac{1}{E_i^\alpha-E_k^\beta}+\frac{1}{E_j^\alpha-E_k^\beta}  \right),  \end{gather}
where Greek indices label different energy eigenspaces (with $\alpha=0$ the ground state subspace), the Latin ones states within subspaces and $E_i^{\alpha}$ are the unperturbed energies.

Since the effective Hamiltonian acts (by definition) wholly within a given subspace $\mathcal{H}_\alpha$, and the perturbation $\hat{V}$ generically may couple different such subspaces, then the matrix elements receive contributions from ``virtual" processes coupling $\mathcal{H}_\alpha$ to some $\mathcal{H}_{\beta\not= \alpha}$ space and back, see Eq. (\ref{sw1}). Since for the solvable model we can describe \emph{every} excited state within a topological sector as arising from an application of a string operator creating quasiparticles to one of the states of $\mathcal{H}_0$, then such ``virtual" processes correspond to \emph{closed contractible loops} of string operators which do not take the system out of $\mathcal{H}_\alpha$, and application of boundary operators, which act within. In particular, in the ground state subspace $\mathcal{H}_0$ all closed contractible loops are trivial (equivalent to identity operators), since they can be shown to factor into a product of either $B_P$ or $\exp \left(i\pi Q_s\right)$ operators enclosed by the loop. Hence the effective edge Hamiltonian is a generic function (dependent on the form of perturbation $\hat{V}$) of all the edge operators only:
\begin{equation}\label{hedge2} \hat{H}_e = f_{\hat{V}}(N_s, R_{ss'},b_s). \end{equation}

In general, the $\hat{H}_e$ can contain terms which couple any number of sites. Since the bulk is gapped, we expect it to be short-ranged (i.e. the amplitude of terms in $\hat{H}_e$ decays exponentially in their range). In addition, $\hat{H}_e$ reflects the symmetry of the problem: if the $U(1)$ symmetry of the bulk is preserved on the edge, it is invariant under the transformation $b_s \mapsto b_s e^{i
	\chi}$, where $\chi$ is an arbitrary phase.

As a concrete example, we consider the following bulk perturbation:
\begin{equation}
\epsilon \hat{V} = U \sum_s n_s^2 - J \sum_{s,s'}\left( U_{ss'} + {H.c.}\right) -h \sum_s \left( b_s^\dagger + H.c. \right), \label{eq:V}
\end{equation}
with coupling constants $U$, $J$, $h$. The first two terms describe hopping and interactions between particles on the lattice sites. These terms spoil the exact solvability of the bulk Hamiltonian (\ref{HBose}); however, since the bulk spectrum is gapped, it remains in the same phase for sufficiently small $\epsilon$. The third term breaks the global $U(1)$ symmetry, Eq.~(\ref{Qel}); in an easy-axis quantum spin liquid, where the $U(1)$ symmetry corresponds to the $z$ component of the total spin, such a term can describe an in-plane applied magnetic field. 

Using Eq.~(\ref{sw1}), we can derive the effective Hamiltonian of the edge in the presence of the bulk perturbation $\epsilon \hat{V}$. 
To first order in $\epsilon$, the $U$ term in Eq.~(\ref{eq:V}) generates an $N_s^2$ terms at the edge. 
This follows from expressing $n_s$ using Eq. (\ref{Qsop}): $n_s = [Q_s - \sum_{s'} n_{ss'}]/2$
(with $Q_s$ replaced by $N_s$ for sites on the edge). We then write $n_{ss'} = [1 - (-1)^{n_{ss'}}]/2$ . The $(-1)^{n_{ss'}}$ operators create flux excitations, since $\{ (-1)^{n_{ss'}}, B_P \} = 0$ [see discussion below Eq.~(\ref{eq:Bp})]. An explicit calculation shows that, projected to the low energy subspace, we can replace $n_s^2$ by $(Q_s-2)^2/2$ in the bulk, and by $(N_s-1)^2/2$ at the edge (up to unimportant constants).

Acting with a hopping ($J$) term creates a pair of excitations in the bulk with energy $2v$, as can be seen from Eq.~(\ref{QOCom}). If this term is applied at one of the bonds at the edge, it creates a single excitation with energy $v$. Acting with this term on the neighboring edge bond annihilates the excitation, and generates the operator $R_{ss'}$ at the edge. Finally, the $b^\dagger_s$ operators in the $h$ term create excitations with an energy $v$ in the bulk, but can act within the ground state subspace at the edge.

We therefore obtain the following form of the effective edge Hamiltonian, $\hat{H}_e$:
\begin{eqnarray}\label{hedge3} \hat{H}_e &=& \frac{U}{2}\sum_s (N_s-1)^2 -t \sum_{\langle ss' \rangle} \left( R_{ss'} + R_{ss'}^{\dagger} \right)  \\ \nonumber &-& h \sum_s \left(  b_s + b^{\dagger}_s\right),  
\end{eqnarray}
where $t = 2J^2/v$. Note that the $h$ term is present only if the perturbation $\hat{V}$ breaks the $U(1)$ symmetry.

Using the fact that $N_s$ are ($\mathbb{Z}$-valued) bosonic degrees of freedom for which $R_{ss'}$ act like hopping operators, we can map the above Hamiltonian to the well-studied Bose-Hubbard (BH) model with an additional $U(1)$-breaking term. Let us denote by $a_s,a_s^{\dagger}$ the anihilation/creation operators for the BH model and let $N_s$ still denote the number operator for the bosons (i.e. $N_s = a_s^{\dagger}a_s$). We can then identify $R_{ss'} = a_sa_{s'}^{\dagger}$ for all $\langle ss' \rangle$ in the $T=1$ topological sector (in the $T=-1$ sector we map $R_{ss'}=-a_sa_{s'}^{\dagger}$ for \emph{one} link $\langle ss'\rangle$ and as before for others). Furthermore, $b_s = a_s^2$, since $b_s$ annihilates a site-boson $n_s$ which contributes to $N_s$ with a factor 2 [see Eq.~(\ref{Qsop})]. 
The effective Hamiltonian can be mapped to:
\begin{gather}\label{hedge4} \hat{H}_{BH\Delta} = -t\sum_{\langle ss' \rangle}a^{\dagger}_sa_{s'} + \frac{U}{2}\sum_s (N_s-1)^2 \\ \nonumber
- h \sum_s \left[ (a_s)^2 + (a_s^\dagger)^2 \right]. \end{gather}

The phases of the edge correspond then to the phases of the above Hamiltonian $\hat{H}_{BH\Delta}$.
For the special case of $U(1)$-conserving perturbation $\hat{V}$ we have $h = 0$, and the Hamiltonian reduces to the familiar Bose-Hubbard model with superfluid (SF) and Mott-insulating (MI) phases on the edge.

We emphasize that in the presence of a nonvanishing $h$, the original $U(1)$ symmetry of the solvable model is broken down to $\mathbb{Z}_2$ symmetry of $\hat{H}_{BH\Delta}$. We thus expect $\hat{H}_{BH\Delta}$ to exhibit two phases: $\mathbb{Z}_2$-symmetric and $\mathbb{Z}_2$-broken. 
In order to obtain the global phase diagram of the edge, we now analyze an effective field theory that corresponds to the model (\ref{hedge4}).

\paragraph*{Field theory of the edge --}

We follow the standard procedure~\cite{Giamarchi} of going to the continuum limit of bose-Hubbard model, Eq.~(\ref{hedge4}). This is done by introducing bosonized dual bosonized fields $\phi(x)$, $\theta(x)$, that are related to the physical operators by $a_s \sim \exp(i\theta)$ and $a_s^\dagger a^{\vphantom{\dagger}}_s \sim \frac{1}{\pi}\nabla\phi + \frac{1}{a} e^{2 i \phi + 2 \pi i \rho x } + h.c.$, where $\rho$ is the average density of the $a_s$ bosons per unit cell. 
From Eq.~(\ref{hedge4}), we see that $\rho=1$.  
The short distance cutoff of the theory, of the order of the lattice constant, is denoted by $a$. The expansion for the physical operators contain extra terms with higher harmonics of $\phi$, which are less relevant than the terms displayed above. The fields $\phi(x)$ and $\theta(x)$ satisfy the commutation relation
\begin{equation}\label{ftbhd2} \left[ \phi(x), \theta(x') \right] = i\pi \Theta(x'-x), \end{equation}
where $\Theta(x)$ is a Heaviside step function. 

The continuum effective Hamiltonian has the following form:
\begin{eqnarray}\label{ftbhd} \hat{H} &=& \int dx \, \frac{v}{2\pi} \left[ K (\nabla \theta)^2 + \frac{1}{K} (\nabla \phi)^2 \right]  \\ &-&
\frac{v \lambda}{a^2} \cos \left( 2\phi \right) - \frac{v \Delta}{a^2} \cos \left( 2\theta \right) \nonumber, \end{eqnarray}
where $v$ is the sound velocity in the superfluid phase, $K$ is the Luttinger parameter, $\lambda$ is a dimensionless coupling constant that characterizes the locking of the bosons to the lattice, and ${\Delta}\propto h$. Eq.~(\ref{ftbhd}) has the usual Sine-Gordon form that arises when bosonizing the bose-Hubbard Hamiltonian at integer filling, and contains also 
the $\Delta$ term that can be understood by inserting the boson operator $a_s \sim \exp(i\theta)$ into the $h$ term in Eq. (\ref{hedge4}).

\begin{figure}[h]
	\centering
	\includegraphics*[width=8cm]{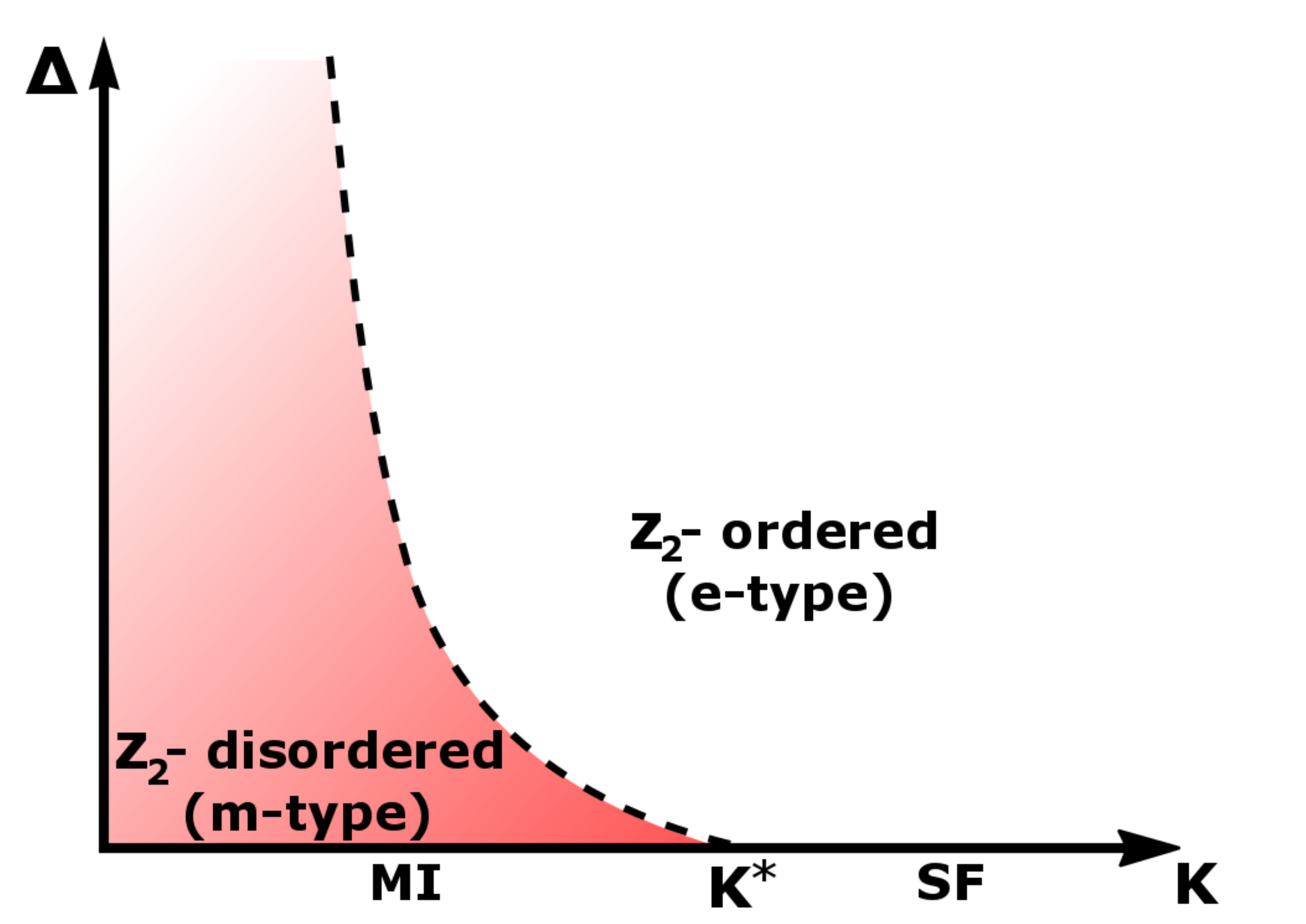}
	\caption{The phase diagram of the edge effective low-energy Hamiltonian, Eq. (\ref{ftbhd})}
	\label{fig:ftpd}
\end{figure}

The leading-order RG equations for $\lambda$ and $\Delta$ are given by:
\begin{eqnarray}\label{rgeq1} \frac{d\lambda}{dt} &=& \left(  2-K \right) \lambda,\nonumber 
\\ \label{rgeq2} \frac{d\Delta}{dt} &=&  \left( 2-\frac{1}{K} \right)\Delta,   \nonumber
\\
\frac{dK}{dt} &=& -\frac{\pi^2 K^2 \lambda^2 }{2} + \frac{\pi^2 \Delta^2 }{2}.
\end{eqnarray}

If the $U(1)$ symmetry is preserved at the edge, i.e. for $\Delta =0$ the $\lambda$-term is relevant for $K < K_{KT} = 2$, marginal for $K=2$ and irrelevant otherwise: those are the (gapped) Mott-insulating and (gappless) superfluid phases of the theory with unbroken $U(1)$ symmetry. The two phases are separated by a Berezinskii-Kosterlitz-Thouless (BKT) transition. Near the transition in the Mott insulator state the gap is of the form $E_{MI} \propto \exp(-C/\sqrt{2-K})$, where $C$ is a non-universal constant.

In the vicinity of the BKT transition, $K \approx 2$, the $U(1)$-breaking perturbation is relevant (see Eq. \ref{rgeq2}) and immediately opens a gap.  We may establish the phase diagram by comparing the magnitudes of the gaps induced by the $\lambda$ and $\Delta$ perturbations. This way we obtain a critical line given by:
\begin{equation}\label{rgcritline}  \Delta_{critical} \sim e^{-\left(2-\frac{1}{K}\right)\frac{C}{\sqrt{2-K}}} \end{equation}

The critical line separates \emph{two} gapped phases: the $\mathbb{Z}_2$-symmetric, smoothly connected to Mott-insulator (characterized by $\langle  \exp(i\theta)\rangle=0$), and the $\mathbb{Z}_2$-broken. The transition between the two phases is of the Ising universality class \cite{PhysRevB.34.6372}. The schematic phase diagram is shown in Fig. \ref{fig:ftpd}.

The identification of the gapped edge phases with the ``e'' and ``m'' topological edges predicted for the $\mathbb{Z}_2$ spin liquid by the Lagrangian subgroup classification \cite{PhysRevX.3.021009} follows from the fact that in the MI phase the charge degree  of freedom is gapped. Therefore there is an energy gap to bringing  the \emph{spinon} excitation of the $\mathbb{Z}_2$ spin liquid (the ``e'' particle) to the edge.  There is no energy penalty for introducing a twist in the boundary conditions, i.e. brining a \emph{vison} (``m'' particle) to the edge, precisely because the charge degrees of freedom are immobile. Those statements taken together are exactly a definition of an ``m''-type edge. Conversely, for the $\mathbb{Z}_2$-ordered phase it can be shown there is a gap to bringing visons to the edge and there is none for bringing spinons, thus it is an ``e''-type edge.

Note that existence of only one gapped edge (the ``m''-type) for $\Delta=0$ and two for non-zero $\Delta$ is fully consistent with the findings of Ref. \cite{Barkeshli722}: though in principle the $\mathbb{Z}_2$ spin liquid can have two topologically distinct edges, only the ``m'' type can be realized when the $U(1)$ symmetry is unbroken. Realizing the ``e''-type edge requires, for instance, placing the system edge in proximity to a ferromagnet or a superconductor.

\section{The entanglement Hamiltonian}\label{sec:ent}
In this section we derive the entanglement Hamiltonian from reduced density matrix $\rho_R$ obtained by tracing out half of degrees of freedom from the groundstate wavefunction and writing it in a thermal form:
\begin{equation}\label{enthamdef1} \rho_R \equiv Tr_L\ket{gs}\bra{gs} \equiv e^{-\hat{H}_{ent}},  \end{equation}
where the trace is over the left part of the system. This defines the entanglement Hamiltonian operator $\hat{H}_{ent}$. We first compute $\hat{H}_{ent}$ for the solvable model and show it has a macroscopically degenerate spectrum, thus equivalent to the unperturbed edge Hamiltonian. Subsequently, we consider the effect of generic perturbations on $\hat{H}_{ent}$ and the correspondence between $\hat{H}_{ent}$ and $\hat{H}_{e}$ derived above. 
The crucial, if a bit technical, step is to introduce a new set of operators which act on the entanglement degrees of freedom at the spatial cut and rewrite the model in their terms, allowing us to carry out the tracing procedure. We refer to the appendices for some of the details.

\subsection{The solvable model} 
\begin{figure}[h]
	\centering
	\includegraphics*[width=9cm]{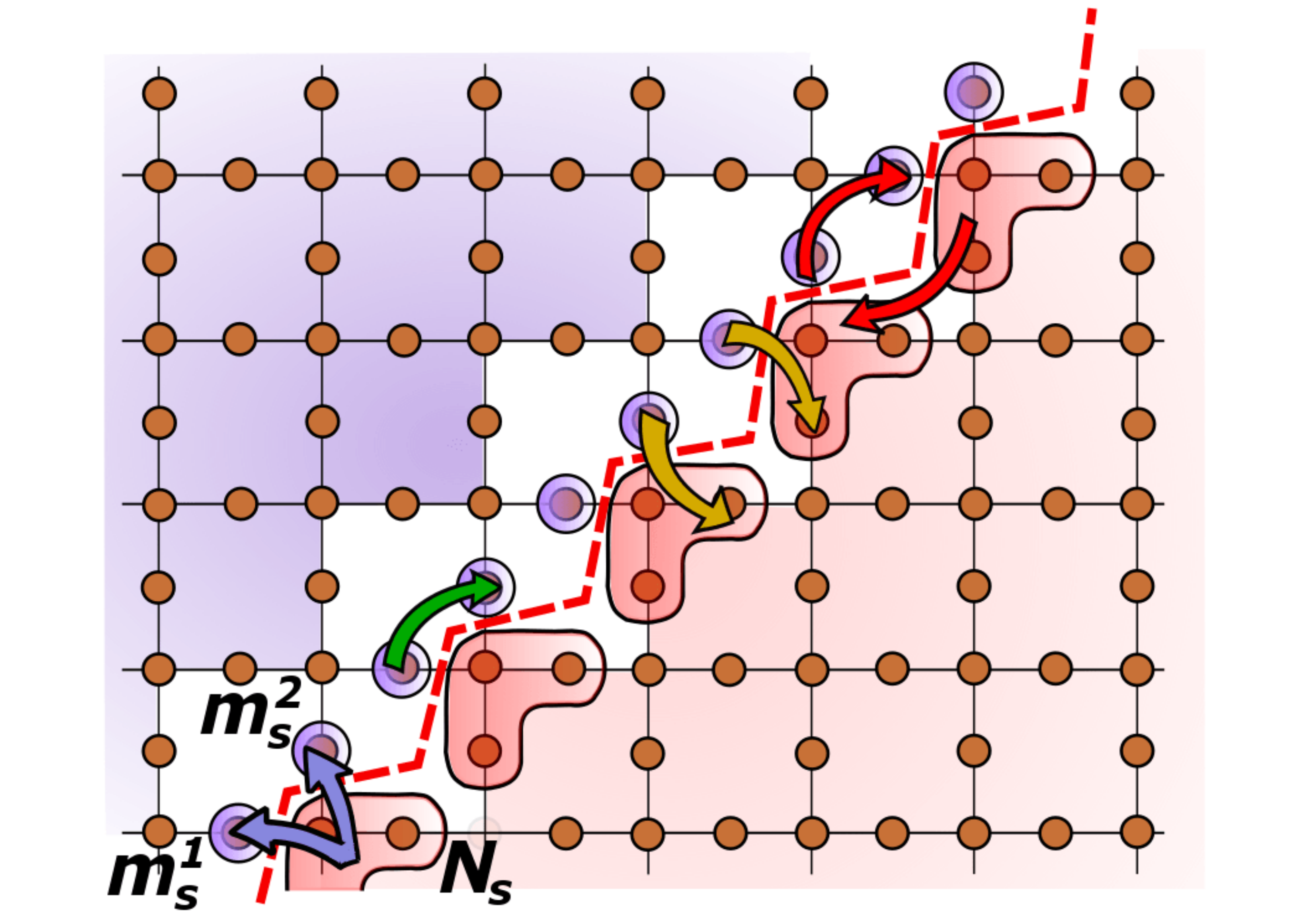}
	\caption{A bipartition of a system on infinite cylinder (periodic vertical direction), the red line denotes the cut. The sites in the purple/red regions to the left/right of the entanglement cut are supports of bulk Hamiltonians $H_{b,L/R}$. The Hamiltonian $H_{LR}$ contains operators whose support includes both ``left" and ``right" sites.. Four types of operators coupling the L/R parts in $H_{LR}$ appearing in last four lines of Eq. (\ref{hlr25}) are schematically depicted using the blue, green, yellow and red arrows. Further explanations in the text.}
	\label{fig:latticem3}
\end{figure}

Consider the system on an infinite cylinder and consider a bipartition of the total Hilbert space into the left and right parts: $\mathcal{H} = \mathcal{H}_L\otimes\mathcal{H}_R$. The cut defining the bipartition is depicted in Fig. \ref{fig:latticem3} by a dashed line.
Note that the left and right edges created by the cut are not related by symmetry.

The entanglement Hamiltonian $\hat{H}_{ent}$ is obtained from the reduced density matrix constructed from the groundstate. In order to obtain a description of the groundstate in a form which will allow for a convenient ``integrating out" of half of the degrees of freedom (without loss of generality: the left part) we write the total system Hamiltonian as follows:
\begin{equation}\label{eq:hpartition1} \hat{H} = \hat{H}_{b,L} + \hat{H}_{b,R}+\hat{H}_{LR},\end{equation}
where $\hat{H}_{b,L/R}$ are the bulk Hamiltonians of the left/right part, defined as in Eq. (\ref{HBose}). The support of $\hat{H}_{b,L/R}$ are all the sites to the left/right of the entanglement cut, denoted by a dashed red line in Fig. \ref{fig:latticem3}. The Hamiltonian $\hat{H}_{LR}$ contains all the terms whose support includes both ``left" and ``right" sites.

The ground state $\ket{gs}$ of $\hat{H}$ is a simultaneous ground state of $\hat{H}_{b,L}$,$\hat{H}_{b,R}$ and $\hat{H}_{LR}$, since they all commute. The ground states of the bulk Hamiltonians $\hat{H}_{b,L/R}$ in each topological sector, i.e. the subspaces $\mathcal{H}_{0,L/R}$, are defined by $Q_s=0$ and $B_P=1$ for all $s$,$P$.  The condition of $\ket{gs}$ being a ground state of $\hat{H}_{LR}$ is more complicated, since $\hat{H}_{LR}$ acts on degrees of freedom on both sides of the cut. In other words $\hat{H}_{LR}$ splits the degeneracy of $\mathcal{H}_{0,L}\otimes\mathcal{H}_{0,R}$.  

Let us first use the ``bulk" description of $\hat{H}_{LR}$ as a sum of the $B_P$ and $Q_s$ operators which include at least one site  from either side of the cut in Fig. \ref{fig:latticem3}. Even though a priori we could have defined topological operators $T_{L/R}$ and $P_{L/R}$  of Eqs. (\ref{topop1},\ref{topop2}) independently for the L/R subsystems, for the groundstate $\ket{gs}$ the left and right part have to be in the same topological sector. This follows from the fact that the product of topological operators $T_LT_R=\prod_{P\in supp(H_{LR})}B_P=1$ since the ground state of $\hat{H}_{LR}$ has $B_P=1$ for all plaquettes in its support. Analogous argument shows $P_L=P_R$. There are thus four topological sectors also for the overall ground state $\ket{gs}$ and without loss of generality we can label them by eigenvalues of $T_R,P_R$: $\ket{gs;t,p}$. In what follows we shall implicitly assume $T_R =1$. The other sector, as mentioned before, corresponds to a change of sign of hopping amplitude of the $N_s$ bosons along the edge \emph{on a single link}.

Since some of the degrees of freedom which the $B_P$ and $Q_s$ operators of $\hat{H}_{LR}$ act on will be integrated out, it is now convenient to write $\hat{H}_{LR}$ explicitly using operators which do respect the partition -- the edge operators we defined earlier will come in handy. Note, however, that the edges on both sides of the cut are not equivalent and our description of the edge operators in Section \ref{sec:lattice} applied to the R-subsystem only. We thus introduce the edge operators for the L-subsystem: to this end we first split the $Q_s$ operator (see also Fig. \ref{fig:latticem3}):
\begin{equation}\label{leftedge1} Q_s = m_{s}^{1} + m_{s}^{2} + N_s ,\end{equation}
where $N_s$ is the R-edge operator we defined before, and $m_s^{1,2}$ are by definition the two remaining parts of $Q_s$ on the L-side (which are just the original $\mathbb{Z}_2$-boson variables $n_{ss'}$).

In exact analogy with the R-edge we can also construct edge operators commuting with the left bulk Hamiltonian $H_{b,L}$ from restrictions of the plaquette operators $B_P$ to the left side of the cut. Those restrictions are defined in Appendix A, it is however more convenient to introduce the bosonic creation/anihilation operators $a_{s,1/2,L}$ and $a_{s,1/2,L}^{\dagger}$ for the $m_s^{1,2}$ (which again are equivalent to the original $b_{ss'}$, $b_{ss'}^\dagger$ operators on appropriate links) and to write the $\hat{H}_{LR}$ Hamiltonian directly in their terms.

As shown in Appendix B, the $\hat{H}_{LR}$ Hamiltonian written in the Hilbert space of edge degrees of freedom using the bosonic variables $a_{s,L/R}$ on both sides is given by:
\begin{eqnarray}\nonumber \hat{H}_{LR} &=& v\sum_{s\in\partial}\left( N_s+m_s^1 + m_s^2 \right)^2 + \\ \nonumber
&-& \frac{u}{2}\sum_{\langle s,s+1\rangle \in \partial} \left( a_{s,2,L}a_{s+1,1,L}^\dagger\right) \left( a_{s,R}^\dagger a_{s+1,R}\right) \\ \nonumber
&-& \frac{u}{2}\sum_{\langle s,s+1\rangle \in \partial} \left( a_{s,2,L}a_{s+1,1,L}\right) \left( a_{s,R}^\dagger a_{s+1,R}^\dagger \right) \\ \nonumber
&-& \frac{u}{2}\sum_{s\in\partial}\left(a_{s,1,L}a_{s,2,L}\right)\left(a_{s,R}^\dagger\right)^2 \\  \label{hlr25}
&-& \frac{u}{2}\sum_{s\in\partial} a_{s,1,L}a_{s,2,L}^\dagger +  H.c.\end{eqnarray}
The action of terms in the last four lines of Eq. (\ref{hlr25}) is depicted in Fig. \ref{fig:latticem3} using red, yellow, blue and green arrows, respectively.

Though this Hamiltonian looks rather daunting, it is in fact not difficult to write down the ground state explicitly. We use the fact that the variables $(m_s^1,m_s^2)$ are allowed to take values in $\{(0,0),(0,1),(1,0),(1,1) \}$ only, and that none of the terms in $\hat{H}_{LR}$ changes the parity of the total sum of $m_s^{1,2}$ (or, equivalently, of $N_s$; This follows from parity being a topological operator, i.e. from Eq. \ref{topop4}). Let us denote by $\ket{0_L}\ket{0_R}$ the state defined by $m_s^1=m_s^{2}=N_s=0$ for all $s$. It is a ground state of the charging part of $H_{LR}$ and it is parity-even. We can analogously write down the state $\ket{1_L}\ket{1_R}$ defined by the condition $m_s^1=m_s^{2}=N_s=0$ \emph{except} at one chosen site $t$, where we have $m_t^1 = -N_t = 1$. This state is parity-odd. The ground states of the full $\hat{H}_{LR}$ with definite parity $p$ can now be written by fully symmetrizing $\ket{0_L}\ket{0_R}$ and $\ket{1_L}\ket{1_R}$ w.r.t. the terms in the last four lines of Eq. (\ref{hlr25}):
\begin{equation}\label{gssym} \ket{gs;t,p} = \frac{1}{\mathcal{N}_p} \hat{\mathcal{S}} \ket{p_L}\ket{p_R}, \end{equation}
where $p =0,1$, the operator $\hat{\mathcal{S}}$ implements symmetrization, and $\mathcal{N}_p$ is a normalization factor. The key observation is that, by construction, $\ket{gs;t,p}$ is a totally symmetric superposition of all the states satisfying $m_s^1 + m_s^2 = -N_s$ for all sites $s$, and having total parity $p$.

The reduced density matrix in parity sector $p$ is given by $\rho_R^p = \mathrm{Tr}_{L}\ket{gs;t,p}\bra{gs;t,p}$. Since $\ket{gs;t,p}$ is an equal weight superposition of mutually orthogonal states, $\rho_R^p$ is a projector. This immediately implies that the spectrum $\{ w_{\alpha}\}$ of the reduced density matrix and hence also the entanglement spectrum \emph{is flat} -- it is  equivalent to the spectrum of the edge of the unperturbed system $\hat{H}_e\equiv 0$ modulo a constant shift. For this case, we thus find an exact correspondence between the edge and entanglement spectra for the unperturbed system.

The constant $1/\mathcal{N}_p^2$, equal to the flat reduced density matrix eigenvalues $w_\alpha$, is $\frac{1}{\mathcal{N}_p^2} = \left(\frac{4^N}{2}\right)^{-1}$, as shown in Appendix C. Consequently, the entanglement entropy $S_E$ for this bipartition of the system is given by:
\begin{equation}\label{se}S_E = -\sum_{\alpha}w_{\alpha}\log w_{\alpha} = N\log 4-\log 2.  \end{equation}
The result above displays an area-law part proportional to $N$ and a topological entanglement entropy of  $\log 2$ as expected for a $\mathbb{Z}_2$ spin liquid \cite{PhysRevLett.96.110405,PhysRevLett.96.110404}.

\subsection{The perturbed system}
In the previous sections, we computed the entanglement Hamiltonian for the unperturbed system and found that its spectrum is flat. In Section \ref{sec:edge} we also derived what the structure is for the edge Hamiltonian generated by perturbations. We now examine the entanglement Hamiltonian $\hat{H}_{ent}$ in the presence of small perturbations using the method outlined in Ref. \cite{PhysRevB.91.125119}. We show that the effective entanglement Hamiltonian acting on the ``low-entanglement energy'' subspace (i.e., the subspace of entanglement states with a high weight) has an expansion in terms of the edge operators, and it is short ranged. I.e., it has the same structure as the effective Hamiltonian of a physical edge. The coupling constants of the two Hamiltonians, however, are generically different. 

Let the system be described by the following generic Hamiltonian:
\begin{equation}\label{hpertent} \hat{H} = \hat{H}_{b,L} + \hat{H}_{b,R} + \hat{H}_{LR} + \epsilon\left( \hat{V}_{b,L} + \hat{V}_{b,R} + \hat{V}_{LR}.  \right) \end{equation}
For $\epsilon=0$ the (flat) spectra of the reduced density matrix of the right subsystem in the even/odd parity sector contained $2^{2N-1}$ non-zero eigenvalues for a cylinder of circumference $N$. Obviously, for $\epsilon\not= 0$ there will be many more non-zero eigenvalues, since the perturbations mix previously decoupled subspaces. However, there are two distinct classes of such eigenvalues: (i) small deformations of the unperturbed non-vanishing eigenvalues and (ii) small deformations of the previously vanishing eigenvalues. The deformations of (ii) appear with a prefactor of $\epsilon$ or higher and hence the corresponding eigenvalues of the entanglement Hamiltonian go as $-\ln(\epsilon) \rightarrow\infty$. There is, therefore, a well defined notion of high- and low-energy part of the entanglement spectrum. We are interested in the latter.

The unperturbed ground state of the total Hamiltonian, Eq. (\ref{eq:hpartition1}), can be written in a different form:
\begin{gather}\label{hlrgs2} \ket{gs;t,p} = \frac{1}{\mathcal{N}_p} \sum_{\vec{m}\in \{0,1\}^{2N}} \mathbb{P}_p^R\ket{\vec{m}}_L\ket{-\vec{m}}_R  \\ \nonumber
= \frac{1}{\mathcal{N}_P} \sum_{\vec{m}\in \mathbb{Z}_2^{2N}} \sum_{\vec{n}\in \mathbb{Z}^N} \ _R\bra{-\vec{n}} \mathbb{P}_p^R \ket{-\vec{m}}_R\cdot\ket{\vec{m}}_L\ket{-\vec{n}}_R, \end{gather}
where in the first line the summation is over all configurations $\vec{m} \in \{0,1\}^{2N} = \mathbb{Z}_2^{2N} $ of the $(m_s^1,m_s^2)$ variables on the left side of the cut, which in the ground state satisfy $m_s^1+m_s^2=-N_s$. By a slight abuse of notation we denote by $\ket{-\vec{m}}_R$ the unique configuration of $N_s$ satisfying those constraints for a given state of the left variables $\ket{\vec{m}}_L$. The states $\ket{-\vec{m}}_R$, $\ket{\vec{m}}_L$ are  groundstates of the right/left bulk Hamiltonians, i.e. they belong to $\mathcal{H}_{0,L/R}$. The operator $\mathbb{P}_p^R$, acting on the right-side variables only, is a projector onto configurations with a total parity $p$. In the second line a resolution of identity was inserted for the R-side Hilbert space: the additional summation is over all possible integer configurations of the $N_s$ variables, i.e. over $\mathbb{Z}^N$. We also denote by $\left( \mathbb{P}_p^R \right)_{nn'}$ the matrix element $_R\bra{-\vec{n}} \mathbb{P}_p^R \ket{-\vec{n'}}_R$.

The perturbed ground state $\ket{gs^*;t,p}$ of the Hamiltonian in Eq. (\ref{hpertent}) can be written in an analogous fashion:
\begin{align} \ket{gs^*;t,p} &= \frac{1}{\mathcal{N}_p} \sum_{\vec{m},\vec{n}} \left ( \mathbb{P}_p^R - \mathbb{P}_p^R\Lambda\mathbb{P}_p^R \right)_{nm} \ket{\vec{m}}_L\ket{-\vec{n}}_R \nonumber \\ \nonumber
&+ \sum_{\vec{m}}\sum_{\alpha > 0, i}A_{\vec{m};(i,\alpha)}\ket{\vec{m}}_L\ket{i,\alpha}_R\\ \nonumber  &+ \sum_{\vec{n}}\sum_{\alpha > 0, i}B_{\vec{n},(i,\alpha)}\ket{i,\alpha}_L\ket{\vec{n}}_R  \\ \label{hlrgs3a}
&+ \sum_{\alpha > 0, i}\sum_{\beta > 0, j} C_{(i,\alpha);(j,\beta)}\ket{i,\alpha}_L\ket{j,\beta}_R, \end{align}
where $(\mathbb{P}_p^R\Lambda\mathbb{P}_p^R)_{nm}$ describes the correction to the ground state in the low-energy subspace (which is in the same parity sector).
 The $A,B,C$ coefficients describing the contributions of higher energy ($\alpha > 0$) states in the excited bulk subspaces $\mathcal{H}_{\alpha,L/R}$ are of order at least $\epsilon$. Consequently, their contributions to the reduced density matrix $ \rho_R=\mathrm{Tr}_L\ket{gs^*;t,p}\bra{gs^*;t,p}$ come at order at least $\epsilon^2$, or, in other words, to linear order in $\epsilon$ we have:
\begin{equation}\label{redrho1} \rho_R = \frac{1}{\mathcal{N}_p^2} \sum_{\vec{n},\vec{n}'} \left( \mathbb{P}_p^R - 2\mathbb{P}_p^R(\Lambda + \Lambda^\dagger)\mathbb{P}_p^R \right)_{n'n} \ket{-\vec{n}}_R\bra{-\vec{n'}}_R \end{equation}

In the fixed parity $p$ sector the parity projector acts as an identity;  we have then:
\begin{equation}\label{rhoexp1} \rho_R \approx \frac{1}{\mathcal{N}_p^2} e^{-2\mathbb{P}_p^R(\Lambda + \Lambda^\dagger)\mathbb{P}_p^R}\mathbb{P}_p^R, \end{equation}
which allows for the identification of the entanglement Hamiltonian of the perturbed system:
\begin{equation}\label{rhoexp2} \hat{H}_{ent}^* =  2\mathbb{P}_p^R(\Lambda + \Lambda^\dagger)\mathbb{P}_p^R. \end{equation}
In order to derive the entanglement Hamiltonian (to lowest order in perturbations) we thus need to compute the ground state correction $\mathbb{P}_p^R\Lambda\mathbb{P}_p^R$. To this end, following Ref. \cite{PhysRevB.91.125119}, we rewrite the perturbed Hamiltonian Eq. (\ref{hpertent}) in a form which clearly separates, order-by-order in $\epsilon$, terms acting within and between the unperturbed energy eigenspaces $\mathcal{H}_{\alpha,L/R}$, as well as terms which couple the two sides. The full Hamiltonian can be then written as (see Appendix D):
\begin{eqnarray}\nonumber  \hat{H} &\equiv& \hat{H}_{b,L} + \hat{H}_{b,R} + \hat{H}_{LR} + \epsilon\left( \hat{V}_{b,L} + \hat{V}_{b,R} + \hat{V}_{LR} \right) = \\ \nonumber
&=& \hat{H}_{b,L} + \hat{H}_{b,R} + \hat{H}_{LR} + \hat{H}_{edge,L} + \hat{H}_{edge,R} + \\ \label{sw2a} &\ &+ \epsilon \hat{V}_{LR} +  \ldots,\end{eqnarray}
where $\hat{H}_{edge,L}$ and $\hat{H}_{edge,R}$ are the effective low-energy Hamiltonians acting on the left and right degrees of freedom, obtained by a Schrieffer-Wolff transformation with respect to $\epsilon \hat{V}_{b,L}$ and $\epsilon \hat{V}_{b,R}$, respectively (see Appendix D for details).

The omitted terms [shown in Eq. (\ref{sw1a})] do not have matrix elements within $\mathcal{H}_{0,L}\otimes\mathcal{H}_{0,R}$ subspace at order $\epsilon$, as opposed to the edge Hamiltonians which do. Furthermore, $\hat{V}_{LR}$ by definition does not create bulk excitations to the left or right from the cut, nor, being local, can it change the topological sector, hence we may consider only its component $\hat{V}_{0,LR}$, which acts in $\mathcal{H}_{0,L}\otimes\mathcal{H}_{0,R}$.

We now want to identify the correction $\Lambda$ to the ground state, which, by definition  [Eq. (\ref{hlrgs3a})], lives in $\mathcal{H}_{0,L}\otimes\mathcal{H}_{0,R}$. To first order in $\epsilon$ we have:
\begin{equation}\label{pertgs1} \ket{gs^*;t,p} = \ket{gs;t,p} - \sum_{ex} \frac{\bra{ex} \epsilon\hat{V}\ket{gs;t,p}}{E_{ex}-E_{gs}}\ket{ex}, \end{equation}
where $\ket{ex}$ is any excited eigenstate of the unperturbed  Hamiltonian $\hat{H}_{b,L} + \hat{H}_{b,R} + \hat{H}_{LR}$, and $E_{ex}$ is its energy, and $E_{gs}$ is the ground state energy. For the corrections to $\Lambda$, however, $\ket{ex}\in \mathcal{H}_{0,L}\otimes\mathcal{H}_{0,R}$, i.e. $\ket{ex}$ must be an excitation of  $H_{LR}$ only [otherwise it creates bulk excitations, which do not contribute to $\Lambda$, see Eq. (\ref{hlrgs3a})]. By Eq. (\ref{sw2a}) the generic perturbation $\epsilon\hat{V}$, to this order, is either $\hat{H}_{edge,L/R}$ or $\epsilon \hat{V}_{0,LR}$.

In Section \ref{sec:edge} we have  shown that  $\hat{H}_{edge,R}$ is a function of edge operators only. We can write it more explicitly as a series expansion:
\begin{equation}\label{expansionf}\hat{H}_{edge,R}(\{c\}) = \sum_{\vec{s},\vec{\alpha}_N,\vec{\alpha}_R,\vec{\alpha}_b} c^{\vec{\alpha}_N,\vec{\alpha}_R,\vec{\alpha}_b}_{\vec{s}}\cdot N_{\vec{s}}^{\vec{\alpha}_N}R_{\vec{s}}^{\vec{\alpha}_R}b_{\vec{s}}^{\vec{\alpha}_s} ,\end{equation}
where we used the multi-index notation to denote a product of the edge operators acting on sites $\vec{s} = (s_{i_1},s_{i_1+1}\ldots,s_{i_1+N})$. The multi-indices $\vec{\alpha} = (\alpha_{i_1},\ldots,\alpha_{i_1+N})$ specify the power of the edge operator at each site $s_i$ (or on pair of sites $(s_i,s_{i+1})$ in the case of $R$ operators).  The spatial dependence of coefficients $c^{\vec{\alpha}_N,\vec{\alpha}_R,\vec{\alpha}_b}_{\vec{s}}$, whose set we denoted by $\{c\}$, is such as to ensure that the edge Hamiltonian is local; this is because the edge Hamiltonian is generated by local perturbations; this also ensures it does not mix topological sectors.
An analogous description holds for  $\hat{H}_{edge,L}$ and $\hat{V}_{0,LR}$, i.e. they can be expanded in terms of the left and right edge operators acting on $\mathcal{H}_{0,L/R}$, since they cannot create bulk excitations. (Note that the left-edge operators are different from the right ones; they are introduced in Appendix A.) 

In Appendix E we consider the matrix element $\bra{ex} \epsilon\hat{V}\ket{gs; t,p}$ for the allowed perturbations in detail and show first that an edge operator $\hat{E} \in \{R_{ss'},b_s \}$ creates a $\hat{H}_{LR}$ eigenstate, and that the resulting correction to the ground state is given by:

\begin{equation}\label{correction2}\Delta_{ \ket{gs;t,p}} =  - \frac{1}{E_{\hat{E}} - E_{gs}} \hat{E}\ket{gs;t,p},\end{equation}
where $E_{\hat{E}}$ is the energy of the eigenstate created by $\hat{E}$.
The operator $N_s$ does not create an eigenstate of $\hat{H}_{LR}$, but a superposition of flux eigenstates and generates a correction to the ground state of the form:
\begin{equation}\label{nsdecomp3maintext} \Delta_{\ket{gs;t,p}} = - \sum_{k=1}^\infty d_k \left(N_s\right)^k\ket{gs;t,p}.\end{equation}

Thus, from  Eqs. (\ref{correction2},\ref{nsdecomp3maintext}) we conclude that the right-edge Hamiltonian
generates a power series in $N_s$ correction to the ground state $\ket{gs;t,p}$ (see Appendix E) of exactly the same form, albeit with rescaled coupling constants  $\{c\}\rightarrow\{\tilde{c}\}$:
\begin{eqnarray}\nonumber\Delta_{\ket{gs;t,p}} &=&  - \sum_{\vec{s},\vec{\alpha}_N,\vec{\alpha}_R,\vec{\alpha}_b} \tilde{c}^{\vec{\alpha}_N,\vec{\alpha}_R,\vec{\alpha}_b}_{\vec{s}}\cdot N_{\vec{s}}^{\vec{\alpha}_N}R_{\vec{s}}^{\vec{\alpha}_R}b_{\vec{s}}^{\vec{\alpha}_s}\ket{gs;t,p}\\ \label{expansionfcorrmaintext}
&=& \hat{H}_{edge,R}(\{\tilde{c}\})\ \ket{gs;t,p} \end{eqnarray}

Furthermore, any left-edge operator $\hat{E}$ applied to the edge degrees of freedom exposed by the entanglement cut can be expressed exclusively in terms of right-edge operators acting in the right subspace and creating the same eigenstate of $\hat{H}_{LR}$ (see Appendix E). Therefore, the action of   $\hat{V}_{0,LR}$ and $\hat{H}_{edge,L}$ on the ground state $\ket{gs;t,p}$ is expressible in terms of the right edge operators only, i.e. they can be both cast in form of Eq. (\ref{expansionf}) and the ground state correction they generate is of the form of Eq. (\ref{expansionfcorrmaintext}).
The total ground state correction generated, to lowest order, by the perturbation $\epsilon\hat{V}$ has thus the form of Eq. (\ref{expansionfcorrmaintext}). 

Comparing the above result with Eq. (\ref{hlrgs3a}) and Eq. (\ref{rhoexp2}) we conclude that:
\begin{equation}\label{Lambdaedge1}\Lambda = \hat{H}_{edge,R}(\{\tilde{c}\}) ,\end{equation}
where $\{c\}$ are the original coupling constants in Eq.(\ref{expansionf}) and $\{\tilde{c} \}$ are the rescaled ones (also by inclusion of terms from $\hat{H}_{edge,L}$ and $\hat{V}_{0,LR}$, which are expressible using the right-edge operators ). 

Since the coupling constants $\{\tilde{c}\}$ are rescaled in a non-uniform fashion the naive expectation of an exact entanglement spectrum to edge spectrum correspondence cannot hold. This result for our model is in agreement with that of Ref. \cite{PhysRevB.91.125119}. What is more important, however, is that all the terms in the entanglement Hamiltonian $\hat{H}^*_{ent}$ are in one-to-one correspondence with the terms in $\hat{H}_{edge,R}$, i.e. they have the same symmetry properties. We thus argue that the phase diagrams of the edge and the entanglement Hamiltonians are in exact correspondence. 

\section{Conclusions and outlook}\label{sec:co}

In this work, we have studied the phase diagram of either the physical edge or the entanglement Hamiltonian of a solvable $\mathbb{Z}_2$ topologically ordered model with a U$(1)$ symmetry, where the spinon excitations carry a fractional U$(1)$ charge, using an exactly solvable model. Within this model, both the physical spectrum at an edge and the entanglement spectrum are macroscopically degenerate. Upon introducing a small perturbation away from the solvable point, we demonstrate that both the physical edge Hamiltonian and the entanglement Hamiltonian take a generic one-dimensional Bose-Hubbard form, and support the same set of phases, as dictated by the bulk topological order and the symmetry of the problem. As long as the global $U(1)$ symmetry is maintained, the edge may either be gapless or in a gapped ($m$-type) phase; if the $U(1)$ symmetry is broken, either in the bulk or at the edge, a gapped $e$-type edge is possible, as well.

We have also analyzed the nature of the phase transitions between the edge phases. When the U$(1)$ symmetry is explicitly broken on the edge, the $e-$type and $m-$type phases are separated by an 1+1 dimensional Ising transition, as anticipated in Ref.~\cite{Barkeshli722} on field theoretic grounds. If the U$(1)$ symmetry is maintained, the gapless phase and the $m$ phase are separated by a Berezinskii-Kosterlitz-Thouless transition.

These features are expected to be generic to $\mathbb{Z}_2$ topologically ordered quantum spin liquids with fractionally charged spinons. The precise edge Hamiltonian depends on microscopic details; however, the possible edge phases and the nature of the phase transitions between them are determined by the bulk topological order and the global symmetry.

It is interesting to contrast our results with those of Ref.~\cite{PhysRevB.86.014404}, where the entanglement Hamiltonian corresponding to resonating valence bond (RVB) wavefunctions was studied. These are specific model wavefunctions for lattice spin systems, that can support $\mathbb{Z}_2$ topological order. It was found that the entanglement Hamiltonian is naturally written in terms of a spin-$\frac{1}{2}$ hard-core particles, whose number is conserved mod$(2)$, reflecting the fact that these particles carry a $\mathbb{Z}_2$ gauge charge [as in Eq.~(\ref{hedge4}) above.] Topologically, the ground state of the entanglement Hamiltonian found in Ref.~\cite{PhysRevB.86.014404} is in the $m$ phase; however, it has an additional \emph{ferromagnetic} order. (This does not imply that there are ferromagnetic correlations in the physical ground state, since the entanglement Hamiltonian is at a finite temperature; see~\cite{PhysRevLett.113.060501}.) This illustrates the fact that the properties of any \emph{particular} phase of the edge, or of the entanglement Hamiltonian, is not uniquely determined by the bulk; only the set of topologically distinct edge phases is.

\acknowledgements{E. B. was supported by the
Minerva foundation, by a Marie Curie Career Integration
Grant (CIG), by the CRC TR 183 (project B03), and by the European Research Council
(ERC) under the European UnionÕs Horizon 2020 research
and innovation programme (grant agreement No.
639172). M.K-J. gratefully acknowledges financial support from the Swiss National Science Foundation (SNSF)}

\appendix

\section{the left edge operators}\label{sec:appA}

The edge operators for the left subsystem can be constructed explicitly using Eqs. (\ref{Bpop},\ref{Uop}) by simply omitting the bosonic operators $b_s$, $b^{\dagger}_s$ from the definitions if $s$ is not part of the left subsystem. Let us denote by $Y_s$ the restriction of the plaquette which contained one of the edge sites $s$ and $Y_{ss'}$ the restriction of a plaquette which contained two neighbouring edge sites $s,s'$ (see Fig. \ref{fig:latticem3}).
We can write the action of those edge operators on the $m_s^{1,2}$ degrees of freedom as using the bosonic operators $a_{s,1/2,L}$, $a_{s,1/2,L}^{\dagger}$:
\begin{eqnarray}\label{leftedge2} Y_s &=& \left( a_{s,1,L} + a_{s,1,L}^\dagger \right) \left(a_{s,2,L} + a_{s,2,L}^\dagger\right),  \\
\label{leftedge3} Y_{ss'} &=& \left( a_{s,2,L}+a_{s,2,L}^{\dagger} \right) \left( a_{s',1,L}+a_{s',1,L}^{\dagger} \right).\end{eqnarray}
The two types of operators originate from plaquette operators in $\hat{H}_{LR}$, i.e. the left and right diagonal rows of plaquettes comprising the white border region around the entanglement cut in Fig. \ref{fig:latticem3}.

Since $Y_s$, $Y_{ss'}$ do not change the eigenvalue of $m_s^{1,2}$ in a definite way, it will be also be convenient to change basis and define auxiliary ``hopping" and ``pairing" operators $R^L_{s,c,s',c'}$ and $Z^L_{s,c,s',c'}$, which do:
\begin{gather}\label{eq:rrop} R^L_{s,c,s',c'} = a_{s,c,L}a_{s',c',L}^\dagger,  \\
\label{eq:zlop} Z^L_{s,c,s',c'} = a_{s,c,L}a_{s',c',L},  \end{gather}
with $s,s'$ sites and $c=1,2$. The $Y$ operators are superpositions of appropriate $Z$ and $R$ operators. We can also define the pairing operator $Z_{ss'}^R$ for the right side analogously.

\section{the $\hat{H}_{LR}$ Hamiltonian in terms of edge operators}\label{sec:appB}

Since the  Hamiltonian $\hat{H}_{LR} = v\sum_{s\in supp(H_{LR})} Q_s^2 - u/2 \sum_{P\in supp(H_{LR})} \left( B_P + B_P^\dagger \right)$ fully commutes with the left and right bulk, we want to write its nontrivial action on the left and right edge degrees of freedom around the cut. The $Q_s$ operator is rewritten using Eq. (\ref{leftedge1}). Since $[B_P,Q_s]=0$ then the action of any plaquette $B_P$ belonging to $\hat{H}_{LR}$ on $m_s^{1,2}$ and $N_s$ must obey:
\begin{equation}\label{appendixeq1}\Delta (m_s^1+m_s^2) = -\Delta  N_s, \end{equation}
where $\Delta$ denotes change of $m_s^{1,2}$ and $N_s$ eigenvalue upon applying $B_P$ to a given state.
Using the fact the the restrictions of $B_P$ to the right subsystem yield the right edge operators $R_{ss'}$ and $b_s$ which change the $N_s$ eigenvalue in a well-defined fashion we can rewrite $B_P$ using $R$ and $b$ operators on the right and $Z$, $R$ on the left, such that each term satisfies the constraint Eq. (\ref{appendixeq1}).

Thus the Hamiltonian $\hat{H}_{LR}$ can equally well be written as:
\begin{eqnarray}\label{hlr2zr} \hat{H}_{LR} &=& v\sum_{s\in\partial}\left( N_s+m_s^1 + m_s^2 \right)^2 \\ \label{hlr2zr2}
&-& \frac{u}{2}\sum_{\langle s,s+1\rangle \in \partial}  R_{s,2,s+1,1}^L \left( R_{s,s+1}^R\right)^\dagger \\ \label{hlr2zr3}
&-& \frac{u}{2}\sum_{\langle s,s+1\rangle \in \partial} Z_{s,2,s+1,1}^L \left( Z_{s,s+1}^R \right)^\dagger \\ \label{hlr2zr4}
&-&\frac{u}{2} \ \ \ \sum_{s\in\partial} Z_{s,1,s,2}^L b_s^\dagger \\  \label{hlr2zr5}
&-&\frac{u}{2}\ \ \  \sum_{s\in\partial} R_{s,1,s,2}^L +  H.c.\end{eqnarray}

Using explicit definitions of $Z$ and $R$ operators Eqs. (\ref{eq:rrop},\ref{eq:zlop}) we obtain the Hamiltonian in Eq. (\ref{hlr25}).

\section{entanglement spectrum of the unperturbed system}\label{sec:appC}
The value of the constant, $1/\mathcal{N}_p^2$, can be calculated explicitly by counting how many terms contribute to $\ket{gs;t,p}$ in Eq. (\ref{gssym}) for a system edge of length $N$ sites, and the counting itself is made easy by the fact that  the variables $(m_s^1,m_s^2)$, which fully constrain their partner $N_s$, only assume  values in $\{(0,0),(0,1),(1,0),(1,1) \}$, two of which are of even-parity and two of odd-parity. Denoting by $E_N$, $O_N$ the number of terms in Eq. (\ref{gssym}) for even/odd parity states of length $N$, and considering the addition of one site to the chain of length $N-1$ we obtain the following recursion relation:
\begin{equation}\label{recrel1} \vvectwo{E_N}{O_N}=\mtwo{2}{2}{2}{2}\vvectwo{E_{N-1}}{O_{N_1}}, \end{equation}
with the initial condition $(E_1,O_1)^T =(2,2)^T$. Solving the recursion we obtain:
\begin{equation}\label{recrel2} \vvectwo{E_N}{O_N} = \vvectwo{\frac{4^N}{2}}{\frac{4^N}{2}}. \end{equation}
Thus the reduced density matrix eigenvalues for both even/odd parities are given by:
$w_\alpha = \frac{1}{\mathcal{N}_p^2}= \left(\frac{4^N}{2}\right)^{-1}$.

\section{Schrieffer-Wolff transformation of the Hamiltonian}

To obtain the groundstate correction $\mathbb{P}_p^R\Lambda\mathbb{P}_p^R$ in Eq. \ref{rhoexp2}, we make use of the fact that the Schrieffer-Wolff transformation generating the effective R/L-edge Hamiltonians may be written as a unitary rotation with $U=e^{-S_\sigma}$, with $\sigma=R/L$ and $S_\sigma$ off-diagonal in the energy subspaces. Let $\mathbb{P}_{\alpha,\sigma}$ be the projector to the left/right bulk Hilbert subspace of energy $\alpha$, such that $\sum_{\alpha\geq 0} \alpha\mathbb{P}_{\alpha,\sigma} = H_{b,\sigma}$, then:

\begin{eqnarray}\label{sw1a} \hat{H}_{\sigma} &=& e^{-S_\sigma}e^{S_\sigma}\left( \hat{H}_{b,\sigma} + \hat{V}_{b\sigma}\right)e^{-S_\sigma}e^{S_\sigma}  \\ \nonumber &=&e^{-S_\sigma}\left( \mathbb{P}_{1,\sigma} + 2\mathbb{P}_{2,\sigma} + \ldots + \hat{H}_{edge,\sigma} + \hat{H}_{1,\sigma} +\ldots \right)e^{S_\sigma} \\ \nonumber &=& \sum_{\alpha \geq 0}\alpha\mathbb{P}_{\alpha,\sigma} + \hat{H}_{edge,\sigma} + \sum_{\alpha \geq 1} \hat{H}_{\alpha,\sigma} \\ \nonumber &+& \sum_{\alpha\geq1} [-S_\sigma, \alpha \mathbb{P}_{\alpha,\sigma}]
- [S_\sigma,\hat{H}_{edge,\sigma}] - \sum_{\alpha\geq 1} [S_\sigma,\hat{H}_{\alpha,\sigma}]  \\ \nonumber
&+& \frac{1}{2}\sum_{\alpha\geq 1} [S_\sigma,[S_\sigma,\alpha\mathbb{P}_{\alpha,\sigma}]] + \ldots. \end{eqnarray}

The first equality is essentially a definition of the Schrieffer-Wolff transformation: the rotated Hamiltonian is written as a sum of projectors to the original subspaces $\mathcal{H}_\alpha$ with bulk energy $\alpha$ and effective Hamiltonians $\hat{H}_{\alpha,\sigma}$ generated by perturbations acting within those subspaces and endowing them with a dispersion. The effective Hamiltonian acting in the space $\mathcal{H}_{0,\sigma}$ is $\hat{H}_{edge,\sigma}$. The second equality follows from applying the Campbell-Baker-Hausdorff formula. Since $\sum_{\alpha\geq 0} \alpha\mathbb{P}_{\alpha,\sigma} = H_{b,\sigma}$ we can rewrite the full system Hamiltonian as:
\begin{eqnarray}\nonumber  \hat{H} &\equiv& \hat{H}_{b,L} + \hat{H}_{b,R} + \hat{H}_{LR} + \epsilon\left( \hat{V}_{b,L} + \hat{V}_{b,R} + \hat{V}_{LR} \right) \\ \nonumber
&=& \hat{H}_{b,L} + \hat{H}_{b,R} + \hat{H}_{LR} + \hat{H}_{edge,L} + \hat{H}_{edge,R}  \\  \label{sw2} &\ &+ \epsilon \hat{V}_{LR} +  \ldots,\end{eqnarray}
where the omitted terms (shown in Eq. \ref{sw1a}) do not have matrix elements within $\mathcal{H}_{0,L}\otimes\mathcal{H}_{0,R}$ subspace at order $\epsilon$, as opposed to the edge Hamiltonians which do.  For more detailed treatment we refer the reader to Ref. \cite{PhysRevB.91.125119}.

\section{corrections to ground state}
Let us examine the matrix element $\bra{ex} \epsilon\hat{V}\ket{gs;t,p}$ of Eq. \ref{pertgs1} in more detail. We argued in Section \ref{sec:ent} that $\epsilon\hat{V}$ is necessarily a function of the edge operators, we thus initially assume $\epsilon\hat{V}$ is one of the right edge operators and calculate the correction; the general result follows from analyticity of $f_{\hat{V}_R}$. The analysis is simplified for the $R_{ss'}$ and $b_s$ edge operators, as they create exact eigenstates of $\hat{H}_{LR}$.

Consider first the term $\bra{ex} R_{ss'}\ket{gs;t,p}$: since $R_{ss'} = U_{sx}U_{xs'}$ [see Fig. \ref{fig:latticem2} and Eqs. (\ref{QOCom},\ref{eq:rss1})] it creates an exact eigenstate of $H_{LR}$ with two $Q_s$ excitations at sites $s$ and $s'$. Hence in the sum over excited states of $\hat{H}_{LR}$ in Eq. (\ref{pertgs1}) there is exactly one non-vanishing matrix element for which $\ket{ex} = R_{ss'}\ket{gs;t,p}$ -- that matrix element is identically one and the ground state receives a correction:
\begin{equation}\label{deltaer}\Delta_{\ket{gs;t,p}} = - \frac{1}{E_R - E_{gs}} R_{ss'}\ket{gs;t,p},\end{equation}
where $E_R$ is the energy of the eigenstate created by $R_{ss'}$.
Thus the operator $R_{ss'}$ appearing in $\hat{H}_{edge,R}$ is reproduced as a correction to the ground state, but --  crucially  -- only up to the energy factor scaling. 

The operator $b_s$ also creates an exact eigenstate of $\hat{H}_{LR}$ with two units of $Q_s$ charge at site $s$; by the same reasoning it is reproduced as a correction to the ground state $b_s\ket{gs;t,p}$, albeit with a \emph{different} scaling $1/(E_b-E_{gs})$. 

For the $N_s$ operator the analysis is slightly more involved, since $N_s$ does not create an eigenstate of $\hat{H}_{LR}$. Instead the string operator $(-1)^{N_s}$ does: it creates an eigenstate of $\hat{H}_{LR}$ with two flux excitations on plaquettes in the support of $\hat{H}_{LR}$ whose link variables $n_{ss'}$ belong to $N_s$ [see Fig.(\ref{fig:latticem2})]. This operator is, however, a power series in $N_s$. Furthermore, while $N_s$ does not create an eigenstate of $\hat{H}_{LR}$, it is clear that it creates a superposition of pure flux eigenstates on three plaquettes sharing either side or corner with $N_s$; this is a consequence of $N_s$ not commuting with the three $B_P$ operators on those plaquettes (and commuting with every other operator in $\hat{H}_{LR}$). Every such flux eigenstate can be created by application of either $A=(-1)^{N_s}$ (for flux states without flux on the plaquette sharing corner with $N_s$) or $B=(-1)^{m_s^1}$ and the product $AB$ (for the states with excitation on the corner plaquette ) to the ground state $\ket{gs;t,p}$, as discussed in Section \ref{sec:lattice}. More explicitly, we have:

\begin{equation}\label{nsdecomp1} N_s = d_{id}\cdot 1 + d_AA + d_BB+ d_{AB}AB, \end{equation}
where $d_{id}$, $d_A$, $d_B$, $d_{AB}$ are complex coefficients. Therefore, for the $N_s$ operator the ground state correction reads:
\begin{equation}\label{nsdecomp2} \Delta_{\ket{gs;t,p}} = - \sum_{\hat{O} \in \{A,B,AB\}}\frac{d_{\hat{O}}}{E_{\hat{O}} - E_{gs}} \hat{O}\ket{gs;t,p}.\end{equation}
As we argued $A =(-1)^{N_s} $ is a power series in $N_s$. For $B = (-1)^{m_s^1}$ this is less evident, since it is written as a function of a \emph{left} edge operator $m_s^1$.  Using Eq. \ref{nsdecomp1}, however, as well as expansion of $A$, it is possible to find an expansion of $B$ as a power series in $N_s$. Using this expansion in Eq. (\ref{nsdecomp2}) we arrive at the expression for the correction to $\ket{gs;t,p}$ due to $N_s$:
\begin{equation}\label{nsdecomp3} \Delta_{\ket{gs;t,p}} = - \sum_{k=1}^\infty d_k \left(N_s\right)^k\ket{gs;t,p}.\end{equation}
The coefficients $d_k$ are functions of eigenstate energies $E_A$, $E_B$, $E_{AB}$. The corrections due to higher powers of $N_s$ in the perturbation $\epsilon \hat{V}$ are obtained analogously and have a similar form.  

The above results can be summed up in the following fashion: the $R_{ss'}$ and $b_s$ edge operators are reproduced, exactly up to an energy dependent scaling factor, as corrections to the ground state. The edge operator $N_s$ generates higher powers of $N_s$ in the correction to the ground state. Note, however, that no other operators are generated,  nor is there any mixing between different edge operators. Thus, a generic right edge Hamiltonian of the form introduced  in Eq. (\ref{expansionf}):
\begin{equation}\label{expansionfapp}\hat{H}_{edge,R}(\{c\}) = \sum_{\vec{s},\vec{\alpha}_N,\vec{\alpha}_R,\vec{\alpha}_b} c^{\vec{\alpha}_N,\vec{\alpha}_R,\vec{\alpha}_b}_{\vec{s}}\cdot N_{\vec{s}}^{\vec{\alpha}_N}R_{\vec{s}}^{\vec{\alpha}_R}b_{\vec{s}}^{\vec{\alpha}_s}\end{equation}
generates a correction to the ground state $\ket{gs;t,p}$ of exactly the same form, albeit with changed coefficients $\{c\}\rightarrow\{\tilde{c}\}$ [obtainable via Eqs. (\ref{deltaer}-\ref{nsdecomp3})]:
\begin{eqnarray}\nonumber\Delta_{\ket{gs;t,p}} &=&  - \sum_{\vec{s},\vec{\alpha}_N,\vec{\alpha}_R,\vec{\alpha}_b} \tilde{c}^{\vec{\alpha}_N,\vec{\alpha}_R,\vec{\alpha}_b}_{\vec{s}}\cdot N_{\vec{s}}^{\vec{\alpha}_N}R_{\vec{s}}^{\vec{\alpha}_R}b_{\vec{s}}^{\vec{\alpha}_s}\ket{gs;t,p}\\ \label{expansionfcorr}
&=& \hat{H}_{edge,R}(\{\tilde{c}\})\ \ket{gs;t,p} \end{eqnarray}

The correction to the ground state we derived in Eq. (\ref{expansionfcorr}) was due to a right-edge Hamiltonian, expressed in terms of right-edge operators. However, in the process, we expressed the effect of action of a \emph{left} edge operator $B = (-1)^{m_s^1}$ in terms of right-edge operators $N_s$. This can be done methodically for any left-edge operator, as we now argue.

Consider a left-edge operator as defined in Appendix A and recall we are interested, via Eq. (\ref{pertgs1}), in perturbations creating excitations of the $\hat{H}_{LR}$ part of the Hamiltonian
(thus staying within the $\mathcal{H}_{0,L}\otimes\mathcal{H}_{0,R} $ subspace). Since the left and right edge operators are paired up in the various terms in $\hat{H}_{LR}$ in Eqs. (\ref{hlr2zr}-\ref{hlr2zr5}), the action of any left-edge operator on the ground state of $\hat{H}_{LR}$ may be expressed in terms of action of the Hermitian conjugate of its right-edge partner in $\hat{H}_{LR}$. For example, from Eq. (\ref{hlr2zr4}) we have $Z^L_{s,1,s,2}\ket{gs;t,p} = b_s\ket{gs;t,p}$. Any left-edge operator can thus be mapped to a right-edge one, creating the same eigenstate of $\hat{H}_{LR}$. 

This left-to-right mapping allows to rewrite the left-edge Hamiltonian $\hat{H}_{edge,L}$, as well as $\hat{V}_{0,LR}$, a priori expressed also in terms of the left-edge operators, into a power series in terms of the right-edge operators only, analogous to Eq. (\ref{expansionfapp}).
Therefore, the correction to $\ket{gs;t,p}$ due to the action of $\hat{H}_{edge,L}$ and $\hat{V}_{0,LR}$ can be written down in terms of right-edge operators entirely, and it has the form of Eq. (\ref{expansionfcorr}). Consequently the full correction to the ground state resulting from the action of $\hat{H}_{edge,R}$ and $\hat{H}_{edge,L}$ and $\hat{V}_{0,LR}$ has this form, with appropriate coefficients $\{\tilde{c}\}$.

\bibliography{ee}

\end{document}